# DATA ANALYSIS OF WIRELESS NETWORKS USING CLASSIFICATION TECHNIQUES

Daniel Rosa Canêdo[1,2] and Alexandre Ricardo Soares Romariz[1]

[1]Department of Electrical Engineering, University of Brasília, Brasília, Brazil
daniel.canedo@ifg.edu.br , alromariz@gmail.com
[2]Federal Institute of Goiás, Luziânia, Brazil
daniel.canedo@ifg.edu.br

## ABSTRACT

*In the last decade, there has been a great technological advance in the infrastructure of mobile technologies. The increase in the use of wireless local area networks and the use of satellite services are also noticed. The high utilization rate of mobile devices for various purposes makes clear the need to track wireless networks to ensure the integrity and confidentiality of the information transmitted. Therefore, it is necessary to quickly and efficiently identify the normal and abnormal traffic of such networks, so that administrators can take action. This work aims to analyze classification techniques in relation to data from Wireless Networks, using some classes of anomalies pre-established according to some defined criteria of the MAC layer. For data analysis, WEKA Data Mining software (Waikato Environment for Knowledge Analysis) is used. The classification algorithms present a success rate in the classification of viable data, being indicated in the use of intrusion detection systems for wireless networks.*

## KEYWORDS

*Wireless Networks, Classification Thecniques, Weka*

## 1. INTRODUCTION

Over the past decade a great technological advance was seen, especially regarding mobile technologies and its infrastructure. The increase in the use of wireless local area networks and also the use of services from satellites, both in organizational and residential environments, is identified. This allows information to be created, transmitted and accessed faster and anywhere at any time by simply having access to the mobile network infrastructure. According to Anatel (Telecommunication National Agency), in January/2016 Brazil registered 257.248 million active lines in mobile telephony, with pre-paid accesses corresponding to 71.45% (183.80 million) of total accesses, while postpaid accesses correspond to 28.55% (73.45 million).

The consequence of this scenario is perceived when the use of computational devices used by both individuals and companies are verified. This scenario can be verified through the research conducted by IDC Brasil, which states that in the last quarter of 2014 Brazil had 1,637 million computers, of which 600 thousand are desktops and 1,037 million are notebooks. An unpublished survey by the Brazilian Institute of Geography and Statistics (IBGE) reveals that 57.3% of homes access the internet through cell phones and tablets in 2013.

People are getting used to technologies such as smartphones and tablets with Internet access. Most of these devices are equipped with capabilities based on the IEEE 802.11 standard. Using these wireless networks, users are often able to gain access to the Internet much cheaper than using cellular networks.

Currently these mobile devices basically act as a small computer, being possible to perform all actions, among others commonly performed on a Personal Computer. Some of these actions are: sending of E-mail to any computational device; use of an operating system; video viewing; execution of Web Systems; content servers; financial transactions; online shopping.

These mobile devices are also part of Wireless Networks as well as wireless actuators offering communication technologies for automation tools built into the Internet of Things in various environments [23].

The high rate of use of mobile devices for various purposes explains the importance of monitoring this infrastructure, since it presents the large-scale transmission of information, which at certain times may be restricted. To the set of this mobile system, determined by both the software and the hardware used, it is relatively fragile with regard to security, mainly due to the characteristic of its transmission medium, but also by the dynamism of access to this system. So there is a need to try to identify the normal and abnormal traffic of these wireless networks so that their administrators can take action.

With increased interconnection between networks, structured and wireless, information security has become a challenge. Networks are subject to various types of attacks that may have internal or external sources, some with the goal of paralyzing services, others with the intention of stealing information and in other cases, just for the amusement of the attackers. In addition, until recently, the networks were restricted to computers, now accept various types of equipment: sensors, smart phones, cell phones, among others. Therefore, security enhancement proposals should consider the technological evolution that is taking place.

The Wireless Networks environment, as well as the environment of Ad Hoc Wireless Networks or Wireless Sensor Networks, has in its characteristic a dynamicity in relation to the composition of the network members, that is, for these types of networks users often enter and leave the network. This feature makes it necessary to manage these environments quickly. This scenario becomes, however, quite vulnerable to attempts to approach the anomalies present in the system as a whole. Anomalies such as *EAPOL Start, Beacon Flood, Deauthentication, RTS Flood* [1][2].

However, the techniques and tools adopted by network managers in the framework of structured computer networks do not always meet these needs in a timely manner. In this sense, the use of intelligent algorithms for classification becomes a great option to minimize these difficulties, in order to identify anomalies more effectively.

The high rate of use of mobile devices for various purposes makes clear the need to monitor this infrastructure, since it presents the large-scale transmission of information, which at certain moments may be confidential. The set of this mobile system, determined by both the software and the hardware used, is relatively fragile regarding security, mainly due to the characteristic of its transmission mean, but also due to dynamic access it. So, there is a need to try to quickly and effectively identify the normal and abnormal traffic of these wireless networks so that administrators can take action. This work aims, from a database of wireless networks [1], to evaluate the classification of these data for some classification techniques. The data is formed by MAC layer information, which will be shown later.

The structure of this article is organized into sections. In section two will be presented some works that have the characteristic of identification of wireless networks traffic using algorithms of learning. In section 3, the theoretical basis for Wireless Networks is presented, while section 4 deals with Classification Techniques. Section 5 will present the methodology of experimentation and results. In Section 6 we present the case studies used to analyze the results.

In section 7 will be performed the quantitative and qualitative analysis of the results. Section 8 presents the conclusion of the work and future work.

## 2. RELATED WORKS

The large increase in the use of mobile computing resources both in public environments, both in private environment has aroused the great use of Ad Hoc Wireless Networks, mainly due to the ease of deployment of these networks. This, in turn, favors the large-scale development of malicious applications in Wireless Networks. It can be said that the number of attacks on Computer Networks, with wireless and structured architecture, has grown in recent years, with the incidents reported in the Brazil exceed 700,000, according to the Center for Studies and Responses to Security Incidents in Brazil [19]. Thus, there is a need to provide resources capable of guaranteeing the minimum authenticity of the services provided by the Computer Networks.

Intrusion Detection Systems are tools that contribute to guarantee the security in the Computer Networks, and its implementation is based on the policy of security of the environment with the objective of keeping active the services made available by the Computer Networks.

In addition, it is necessary to take into account the characteristics of the Ad Hoc Wireless Networks, which make it difficult to monitor the services and components of the Network, since they are constituted by autonomous nodes with mobility and without centralized management. Ad Hoc Wireless Networks rely on direct peer-to-peer communication, which is established without the need for centralized infrastructure. The Ad Hoc Networks are composed of devices that have the cooperative characteristic, being able to establish a direct communication with the devices that are within their reach. In this network there is centralized administration and each device can have the functionalities of station and router. The communication between the stations is called storage-forwarding, that is, the station that wishes to forward a message accesses the transmission medium and forwards the information to the neighboring station, which stores the information until the optimal time to forward the station other than the destination station. In this way, the formation of a multi-hop link between the information source and the destination of the information is identified, making network services such as routing and access control to the medium performed in a distributed way by all the components belonging to Ad Hoc Wireless Network [20].

There are several proposals of Intrusion Detection System in Wireless Networks [2, 5, 7, 21, 22] where the main obstacle is the durability of the energy of the computational resource, being frequently used in these technical proposals of computational intelligence capable of analyzing, learning and identifying anomalies. These proposals are based on the use of classification techniques, either in a single or joint approach, aiming increasingly to better use the mobile computing resource. The result of the use of classification techniques has contributed with the Computer Networks analysts in the choice of security policies with the purpose of nullifying or minimizing the damages caused by the anomalies in Wireless Networks environments.

It is possible to find in the literature some works of Wireless Networks traffic classification, which can be applied in Intrusion Detection Systems. These proposals make use of supervised and unsupervised learning methods. The proposal [2] provides a general approach to the various classification methods, using high-dimensional data and a variable selection technique aiming to reduce computational time and improving the learning rate.

Govindarajan presents a proposal [3] of two classification methods involving multilayer perceptron and Basis function Networks. This work proposes a hybrid architecture involving both classifiers for intrusion detection systems. Ed Wilson presents a proposal [4] of Hybrid

Intrusion Detection System, in which signal processing is performed using the Wavelet transform and then the classification of the anomalies using Artificial Neural Networks.

Ed Wilson[1] proposes the elaboration of a real database of Wireless Network traffic, which will be used in the evaluation of Intrusion Detection Systems (IDS). This data undergoes a pre-processing to later be classified by techniques of standards recognition, such as Artificial Neural Networks and following formatting rules that must be strictly followed.

The proposal [5] uses a combination of selection methods to classify Denial Of Service anomalies in Computer Networks, showing the efficiency of the process selection process for DoS detection.

Vo [6] applies supervised and unsupervised machine learning techniques to predict the time series trend by using the K-Means algorithm to group data with similarity and vector machine to train and test the data.

In [7] the most relevant models for the construction of Intrusion Detection Systems are presented, incorporating machine learning in the scenario of Ad Hoc Wireless Networks. Machine learning methods perform classification approach, association rule mining, Artificial Neural Networks and instance-based learning.

Work [21] also uses unsupervised and supervised classification methods to classify a collection of packet data from the Internet.

Gogoi presents the proposal [22] of a multi-level hybrid intrusion detection method that uses a combination of supervised, unsupervised and discrepant-based methods to improve the efficiency of detecting new and old attacks.

## 3. WIRELESS NETWORKS

The IEEE 802.11 standard defines a structure for the Wireless Local Area Network that covers the physical and link levels present in the reference OSI communication model. For the physical level only, radio frequency (RF) and infrared (IR) transmissions are treated, but other forms of wireless communication such as microwave and visible light can also be considered. For the link level, the access control to the medium is addressed through the definition of the MAC protocol (Medium access Control).

Taking into account the main characteristics of the IEEE 802.11 standard, such as interoperability, low cost, high market demand, reliability of project execution, there is a great growth in the use of Local Area Networks of Wireless Computers, also known as Wireless Networks, in public and private environments. This makes Wireless Networks a priority resource in environments where it is most often possible to access the Internet, whether inside corporations, in homes or in public environments, such as shopping malls, airports and so on [1].

The architecture of Wireless Networks according to the IEEE 802.11 standard is based on the division of the area covered by the Wireless Network into cells, these cells being called BSA (Basic Service Area). The size of the coverage of each BSA will depend exclusively on the characteristics of the environment itself and the power of transmitters and receivers used in the computational devices. The other components of the Wireless Networks architecture are listed below[1]:

I. BSS (*Basic Service Set*): Which is the set of computational devices that communicate by broadcasting (BC) or infrared (IR) within a Basic Service Area;

II. AP (*Access Point*): Specific computational devices, which have the purpose of capturing the transmissions made by computational devices belonging to its BSA (Basic Service Area) and are destined to stations belonging to another Basic Service Area. The Access Point, in turn, will perform the retransmission using a distribution system;

III. Distribution System: Communication infrastructure, which has the purpose of performing the interconnection of several Basic Service Area to allow the construction of networks, which have covers larger than one cell;

IV. ESA (*Extended Service Area*): Service Area that has the purpose of interconnecting several BSAs, through the Distribution System using the Access Point;

V. ESS (*Extended Service Set*): Which is intended to represent a set of computational devices consisting of the union of several BSSs (Basic Service Set) connected by a Distribution System.

The IEEE 802.11 standard also defines a medium access protocol, which is present in a MAC sublayer of the data link level. This protocol is called DFWMAC (*Distributed Foundation Wireless Medium Access Control*), which has two access methods, one of which is a distributed and mandatory feature. The other access method of the DFWMAC protocol is optional, having a centralized feature, and according to the IEEE standard, both the distributed method and the centralized method in the communication system can coexist. The medium access protocol also has the property of treating problems related to computational devices that try to move from one cell to another, a process called roaming. It is also related to the protocol of access to the medium of property to treat problems of lost computational devices, being able to be denominated of hidden node [1].

## 4. CLASSIFICATION TECHNIQUES

Classification is one of the Data Mining techniques that is mainly used to analyze a given dataset and takes each instance of it and assigns this instance to a particular class, thus granting a low error of classification. It is used to extract models that accurately define important data classes within the given dataset. Classification is a two-step process. During first step the model is created by applying classification algorithm on training data set then in second step the extracted model is tested against a predefined test dataset to measure the model trained performance and accuracy. So classification is the process to assign class label from dataset whose class label is unknown.

The dataset evaluation relied on the following classifiers: Bayesian networks, decision tables, Ibk, J48, MLP and NaiveBayes. The main criteria used were the popularity of such classifiers.

Bayesian networks have been used in many approaches to IDS, as in UMER (2017) [8]. These networks are directed acyclic graphics for representing a probability distribution on a set of random variables. Each vertex represents as random variable and each node represents a correlation among the variables [1] [9].

The decision table classifier works representing a set of conditions needed to determine the occurrence of a group of actions by means of a table format [10]. This technique has also been used in IDS approaches [1][11].

The IBk algorithm refers to a way of implementing the kNN (k-nearest neighbor) clustering method, which is used for classification and regression toward finding the closest neighbors of a given instance. In the IBk, three neighbors, the ones closest to the search standard neighbors, are used. This is a relatively simple technique that has been used in IDS approaches as well [1][12].

The J48 algorithm relies on decision tree classifications. By this technique, the classification of a new item depends on the prior creation of a decision tree which uses attributes obtained from the training data. By computing the information gain of each of these attributes, J48 can optimize classification mechanisms in IDS [1][13].

The MLP is an artificial neural network that maps input parameters to proper outputs. It consists of many layers of nodes in a directed graphic. Several IDS approaches have used MLP [1][14].

## 5. METHODOLOGY

This work aims to apply classification techniques to identify anomalies especially in wireless network traffic. As mentioned in the previous section, the techniques adopted for this work are: Bayesian Networks, Decision Tables, Ibk, J48, MLP and NaiveBayes.

In order to achieve the proposed aims, the following activities were performed in accordance with the chronological order of execution.

We use a database with examples of specific anomalies in wireless networks. This base in turn is the final product of the work entitled *A Methodology for building a Dataset to Assess Intrusion Detection Systems in Wireless Networks* [1].

The next step is to perform a pre-processing in the database so that two new databases are obtained. One of the databases is composed of only 10% of the data from the original database and is destined for the test step in the selected algorithms. The other database is composed of 90% of the data from the original database and is destined for the training step of the selected algorithms. Both databases are stored in the Database Manager System named PostgreSQL, and are accessed by the Weka software (*Waikato Environment for Knowledge Analysis*)[15].

Finally, the results of each selected algorithm are analyzed and formatted through tables. In relation to the results, the following information is presented for analysis: Percentage of Classification, relation of correctness and errors.

## 6. CASE STUDY

The case study chosen to analyze the results of the application of classification techniques presented in previous sections uses data from real wireless networks [1] and the data mining software, Weka [15].

### 6.1. Database

The database defined for the execution of this case study is a real collection of network traffic captured in the Wireless architecture. This data, in turn, is obtained by the behavior of users to access different information as well as for the use of the Internet. According to the authors [1], the network traffic obtained by students and employees of the institution in which the experiment was performed was used for this database.

The database chosen for the experimentation of this work made use of two different scenarios. The scenarios discussed have their own configuration and topologies, being a scenario of home environment typical of wireless networks, while the other is a more complex environment, being a corporate environment.

This database is composed of a total of 616,047 records, each record being composed of 16 variables that are characteristics of the wireless network traffic itself. Also in each record of the database is defined a last variable the class to which belongs certain registry, classification is realized taking into account the values of the sixteen variables referring to the obtained wireless network traffic. In this way the data are classified in:

- *Normal*: Acceptable wireless network traffic;

- *EAPOLStart*: Traffic using the Extensible Authentication Protocol (EAP), which aims to perform an authentication method in both the Wired Equivalent Providence (WEP) protocol, both Wi-Fi Protected Access (WPA) protocol, commercial versions for wireless network access;

- *Beacon Flood*: Management type requests, which are intended to transmit millions of invalid Beacons, resulting in the difficulty that a certain Wireless network device will have in identifying a legitimate Access Point [16];

- *Deauthentication*: It also represents management-type requests, which are injected from the Wireless Network. The frames belonging to this anomaly are transmitted as fictitious requests, which requests the deactivation of a device that is authorized in the Wireless Network;

- *RTSFlood*: Also called Request-to-Send Flood is a control-type frame. This anomaly is based on the large-scale transmission of RTS frames or frames for a short period of time [16].

The database for the experimentation process of this work is divided into two distinct bases, in order to meet the requirements of each defined intelligent algorithm. In this way a training database is generated respecting the characteristics of each algorithm, being composed by 554,442 registers, which corresponds to 90% of the complete database. Also, the test database is generated, being composed by 61,604 records that correspond to 10% of the complete database, respecting the characteristic of each algorithm. In order to optimize the experimentation process and to provide better data manipulation, the training and test databases for each defined computational intelligence technique are stored in the PostgreSQL Database Management System.

### 6.2. Dataset Evaluation

The data coming from Wireless Network are evaluated through the classification techniques mentioned in the previous section. To evaluate each of the classification techniques, the error parameters, the percentage of classification and the Kappa coefficient are used, which will be explained later.

The Mean Absolute Error (MAE) is defined as the average of the difference between and computed and measured results. The closer to zero the better the classification is. On the other hand, the Root Mean Square Error (RMSE) is computed as the average of the error square root. A minimum MAE does not imply necessarily in a minimal variation. Thus, it is more effective to use both MAE and RMSE in the evaluations [17].

The MAE and RMSE parameters are a simple way of measuring the effectiveness and efficiency of the classification techniques used, thus they are incentive of more advanced techniques.

The Kappa coefficient, in turn, is initially used by observers in the field of psychology as a measure of agreement-induced [18]. This metric shows the degree of acceptance or agreement among a group of judges. Equation 1 shows the agreement of the Kappa coefficient, with the observed agreement *Po* and the coincidence by chance *Pa*.

$$k = \frac{Po - Pa}{1 - Pa}$$

(1)

The result of *k* = 1 means that the classification was correct, while *k* = 0 indicates that the classification is entirely by chance. However, the best classifiers are those in which the value of *k* is close to one.

As previously shown, the classification techniques to be evaluated for the Wireless Networks database are: Bayesian networks, decision tables, Ibk, J48, MLP and NaiveBayes.

**6.2. Results and Discussion**

The evaluation of the classifiers is performed using the Weka [15] tool, using the set of data obtained from Wireless Network [1] in which they are classified with the following anomalies: EAPOLStart, Beacon Flood, Deauthentication and RTSFlood. This database is composed of 17 variables per record, 16 MAC layer attributes and an identification attribute of the class to which a particular record belongs.

Experimentation with the chosen classification techniques makes use of 90% of training data and 10% of data for testing. The results of the mean and quadratic errors for the same, during the training are shown in Figure 1. These are relatively small, and it can be deduced that the classifiers have good performance for the data set of Wireless Networks.

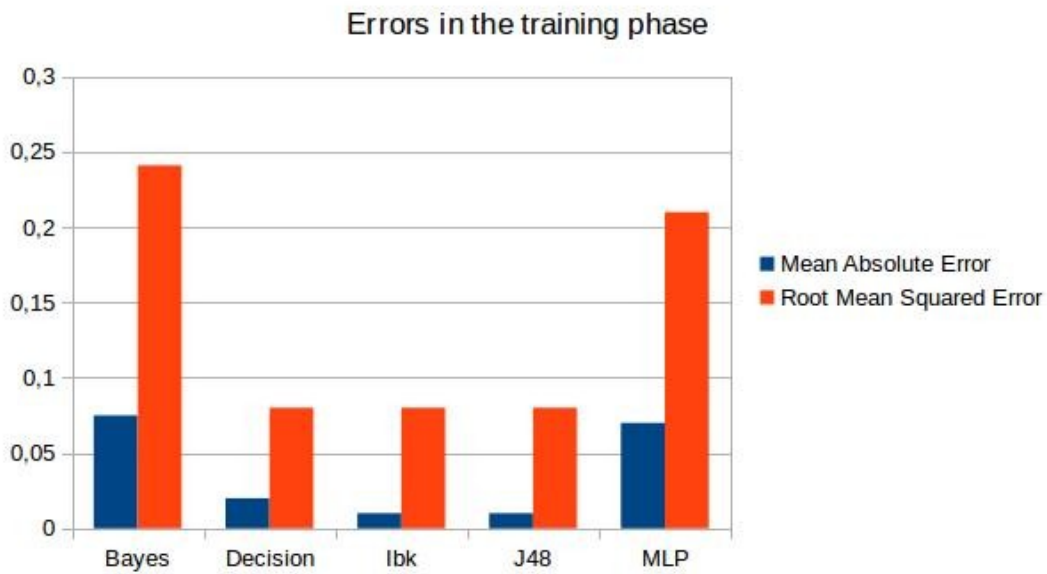

Figure 1.  Errors in the training phase

Table 1 presents the simulation results, after the training of the classification techniques in relation to the Wireless Network data. The values obtained in percentage of correctly classified instances are relevant, being superior to some found in literacy. It should be considered that the proposal is to evaluate the performance of the classification techniques for the application of Wireless Networks data, without customizing them.

Table 1.Results for the testing phase of the data set

| Classification Techniques | Correctly Classified Instances (%) | Incorrectly Classified Instances (%) | Kappa Coefficient |
|---|---|---|---|
| Bayes Network | 76 | 24 | 0,42 |
| Decision Tree | 98 | 2 | 0,91 |
| Ibk | 98 | 2 | 0,91 |
| J48 | 98 | 2 | 0,91 |
| MLP | 75 | 25 | 0,4 |

The evaluation of the data set represents an important research phase in the area of Wireless Networks, as it allows verifying the adequate response of the classification techniques commonly used in Intrusion Detection Systems proposals.

The use of the classification techniques adopted showed good results. The average errors, as shown in Figure 1 are relatively low. It is observed that the absolute mean error as well as the mean square error followed the same trend, proving the actual behavior of the data of Wireless Networks.

Table 1 shows that there is no difference for similar classification algorithms such as Bayes Network and MLP, in which it obtained a rating of 75%, while the other classification algorithms reached 98% of classification with low average errors. Therefore, it is possible to affirm that the use of classification techniques are effective for Wireless Network environments and can be used in Detection and Anomaly Classification Systems for Wireless Networks. It is also noticed that the selection of variables is fundamental for the classification to reach satisfactory levels and optimize the processing of these algorithms.

## 7. CONCLUSIONS AND FUTURE WORK

The results show that the data used to evaluate the classification techniques are viable and can be components in the evaluation of Intrusion Detection Systems in Wireless Networks. However, despite being preformatted with labels, where each record is identified as normal or with some of the predefined anomalies, it becomes valuable because it is collected directly from a Wireless Network.

The errors found in the training phase of the classification algorithms are low, being below 0.25, confirming that the selected classification techniques are adequate and that the data collected from Ad Hoc Wireless Networks are efficient for the analysis of the same ones.

The Kappa Coefficient results follow the same characteristics of the errors in the training phase of the classification algorithms in relation to the correct and incorrectly classified data, thus confirming their integrity.

Future work can be done in several ways: applying these classification techniques to a wireless network, online detection and classification on the network, and comparing with other existing approaches.

**Authors**

**Daniel R. Canêdo** has a degree in Computer Engineering from Pontifícia Universidade Católica de Goiás (2003) and a Master's degree in Electrical Engineering from the University of Brasília (2006). He is currently an exclusive professor at Federal Institute of Goiás - Campus Luziânia. He is currently a PhD student in the Post-Graduate Program in Electronic Systems and Automation Engineering of the Department of Electrical Engineering of the University of Brasília (UnB).

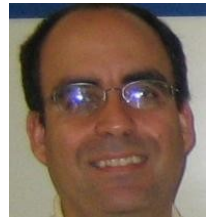

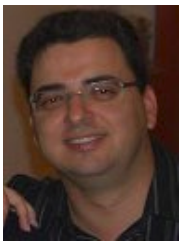

**Alexandre R. Romariz** holds a BS in Electrical Engineering from the University of Brasília (1992), a Master's degree in Electrical Engineering from the State University of Campinas (1995) and a PhD in Electrical Engineering from the University of Colorado at Boulder (2003). He is currently "Professor Associado" at University of Brasilia. He has experience in Computational Intelligence, Integrated Circuits, Optoelectronics and Digital Signal Processing.